\documentclass{article}
\usepackage{hiph-art}
\usepackage{graphicx}
\usepackage{amssymb}

\volnumber{0} \issuenumber{0} \edyear{2005}                              
\frompage{000} \topage{000}                                              
\recrevdate{October 2005}                                                
\def\rightmark{Quark-antiquark production from classical fields}
\def\leftmark{F. Gelis, K. Kajantie and T. Lappi}
 
\title{Production of quark pairs from classical gluon fields}

\authors{ 
{F. Gelis$^1$, K. Kajantie$^2$ and T. Lappi$^{2,3}$ %
\index{Gelis, F.} 
\index{Kajantie, K.} 
\index{Lappi, T.} 
}\\[2.812mm]
{\normalsize
\hspace*{-8pt}$^1$ Service de Physique Th\'eorique, \\
B\^at. 774, CEA/DSM/Saclay, 91191 Gif-sur-Yvette, France\\[0.2ex] 
\hspace*{-8pt}$^2$ Department of Physics, \\
P.O.Box 64, FI-00014 University of Helsinki, Finland\\
\hspace*{-8pt}$^3$ Helsinki Institute of Physics, \\
P.O.Box 64, FI-00014 University of Helsinki, Finland\\
}}
 
\abstract{
We compute by numerical integration of the Dirac equation the number of
quark-antiquark pairs produced in the classical color fields of colliding
ultrarelativistic nuclei. Results for the dependence of the number of quarks
on the strength of the background field, the quark mass and time are presented. 
We also perform several tests of our numerical method.
While the number of $q\bar q$ pairs is parametrically
suppressed in the coupling constant, we find that in this classical field model it
could even be compatible with the thermal ratio to the number of gluons. 
}

 
\makeindex

\newcommand{\qs}{Q_{\mathrm{s}}}
\newcommand{\ra}{R_A}
\newcommand{\nf}{N_\mathrm{f}}
\newcommand{\nc}{N_\mathrm{c}}

\newcommand{\gt}{{\gamma^0}}

\newcommand{\ud}{\mathrm{d}}
\newcommand{\xt}{\mathbf{x}_T}
\newcommand{\yt}{\mathbf{y}_T}

\newcommand{\pt}{{\mathbf{p}_T}}
\newcommand{\qt}{{\mathbf{q}_T}}

\newcommand{\gev}{\textrm{ GeV}}
\newcommand{\fm}{\textrm{ fm}}

\begin{document}
 
\maketitle

\section{Introduction}

Due to large densities, implying large occupation numbers, 
the initial stages of an ultrarelativistic heavy ion collision may be
be dominated by strong classical color fields.
There is a twofold interest in calculating the production of quark--qntiquark pairs
from these classical fields. Firstly, although heavy quark production is 
in principle calculable perturbatively, it would be interesting to understand
whether these strong color fields influence the result. Secondly, being able 
to compute both gluon and quark production in the same framework would give
insight into the chemical equilibration of the system and the consistency
of the assumption of gluon dominance. 
The number of quark pairs present in the early stages of the system
also has observable consequences in the thermal photon and dilepton spectrum.

In this talk we shall present first results \cite{Gelis:2005pb,Gelis:2005gs} of a numerical computation 
of quark antiquark pair production from the classical fields of the 
McLerran-Venugopalan (MV) model. The equivalent calculation, although in the covariant 
gauge unlike the present computation,
 has been carried out analytically to lowest order in the densities of both 
color sources (``pp''-case)  in Ref.~\cite{Gelis:2003vh} and to lowest 
order in one of the sources (``pA''-case) in Ref.~\cite{Fujii:2005vj}.
The corresponding calculation in the Abelian theory \cite{Baltz:1998zb,Baltz:2001dp}, 
of interest for the physics of ultraperipheral collisions,
can be done analytically to all orders in the electrical charge of the
nuclei. 
Quark pair production has also been studied in a related
``CGC''-approach in Refs.~\cite{Kharzeev:2003sk,Tuchin:2004rb} and in a more
general background field in Ref.~\cite{Dietrich:2004eb}.

\begin{figure}[t]
\hfill
\includegraphics[width=0.45\textwidth]{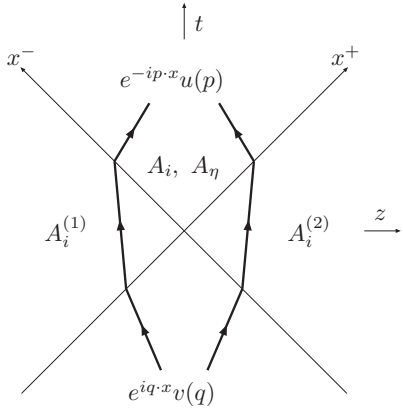}
\hfill
\begin{minipage}[b]{0.49\textwidth}
\caption{
Domains of different time dependences. The fermion amplitude
is a sum of two terms: one interacting first with the left moving
nucleus, then the right moving one, and vice versa.
$A_i^{(1,2)}$ are pure gauges
and $A_i,A_\eta$ is a numerically computed color field.
In the perturbative limit of the Abelian theory
these terms correspond to the $u$ and $t$ channel diagrams in 
Fig.~\protect\ref{fig:diags}.
}\label{fig:spacet}
\end{minipage}
\end{figure}

\section{The numerical calculation}

Our calculation of pair production relies on the numerical calculation of 
the classical background color field, in which
we solve the Dirac equation.

\subsection{The background field}

In the classical field model the background gluon field is obtained from solving
the Yang-Mills equation of motion with the classical color source $J^{\nu}$ given by transverse
color charge distributions of the two nuclei boosted to infinite energy:
\begin{equation}
[D_{\mu},F^{\mu \nu}] = J^{\nu}
=\delta^{\nu +}\rho_{(1)}(\xt)\delta(x^-)
+ \delta^{\nu -}\rho_{(2)}(\xt)\delta(x^+).
\end{equation}
In the MV model \cite{McLerran:1994ni} the color charges are taken to be
random variables with a Gaussian distribution
\begin{equation}
\langle \rho^a(\xt) \rho^b(\yt) \rangle 
= g^2 \mu^2 \delta^{ab}\delta^2(\xt-\yt)
\end{equation}
depending on the coupling $g$  and a phenomenological parameter $\mu$. The combination
$g^2\mu$ is closely related to the saturation scale
$\qs$. Collisions of two ions were first studied analytically using this model
in  Ref.~\cite{Kovner:1995ts} and the way of numerically solving the 
equations of motion in a Hamiltonian formalism in the $A_\tau=0$ gauge
was formulated in Ref.~\cite{Krasnitz:1998ns}.

\subsection{Solving the Dirac equation}

Our method of solving the Dirac equation is explained in more detail 
and the numerics tested in a 1+1-dimensional toy  model
in Ref.~\cite{Gelis:2004jp}. The domains of spacetime involved in the calculation
are illustrated in Fig.~\ref{fig:spacet}.

\begin{figure}[t]
\begin{minipage}[t]{0.49\textwidth}
\begin{center}
\includegraphics[width=0.99\textwidth]{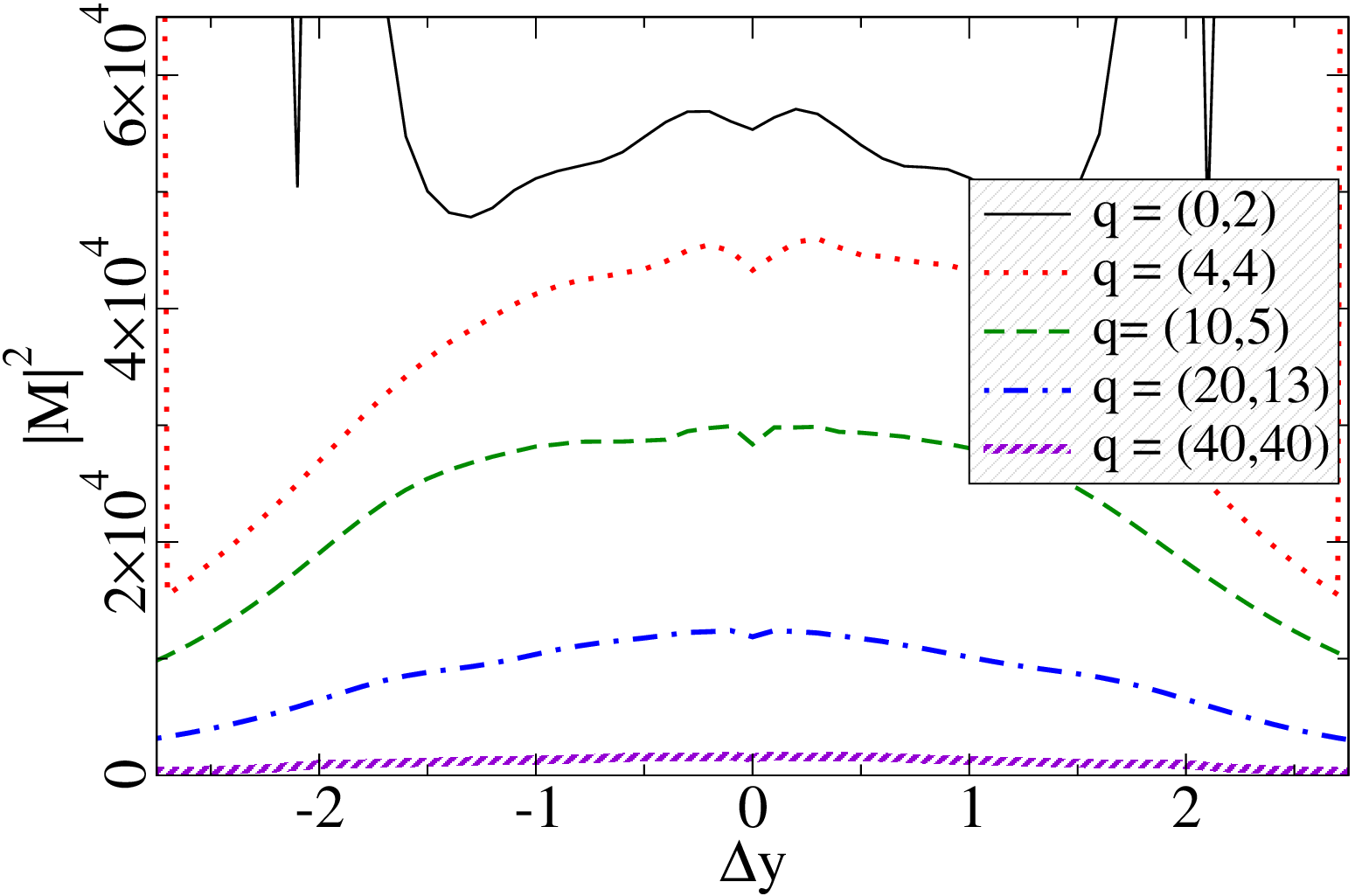}
\end{center}
\vspace{-0.5cm}
\caption{
Amplitude (in lattice units) at $\tau=0.25 \fm$ as a function of $ \Delta y$ 
for different antiquark momenta on a $180^2$-lattice (see text in 
Sec.~\protect\ref{sec:numerics} for the notation).
}\label{fig:ampliy}
\end{minipage}
\hfill
\begin{minipage}[t]{0.49\textwidth}
\begin{center}
\includegraphics[width=0.99\textwidth]{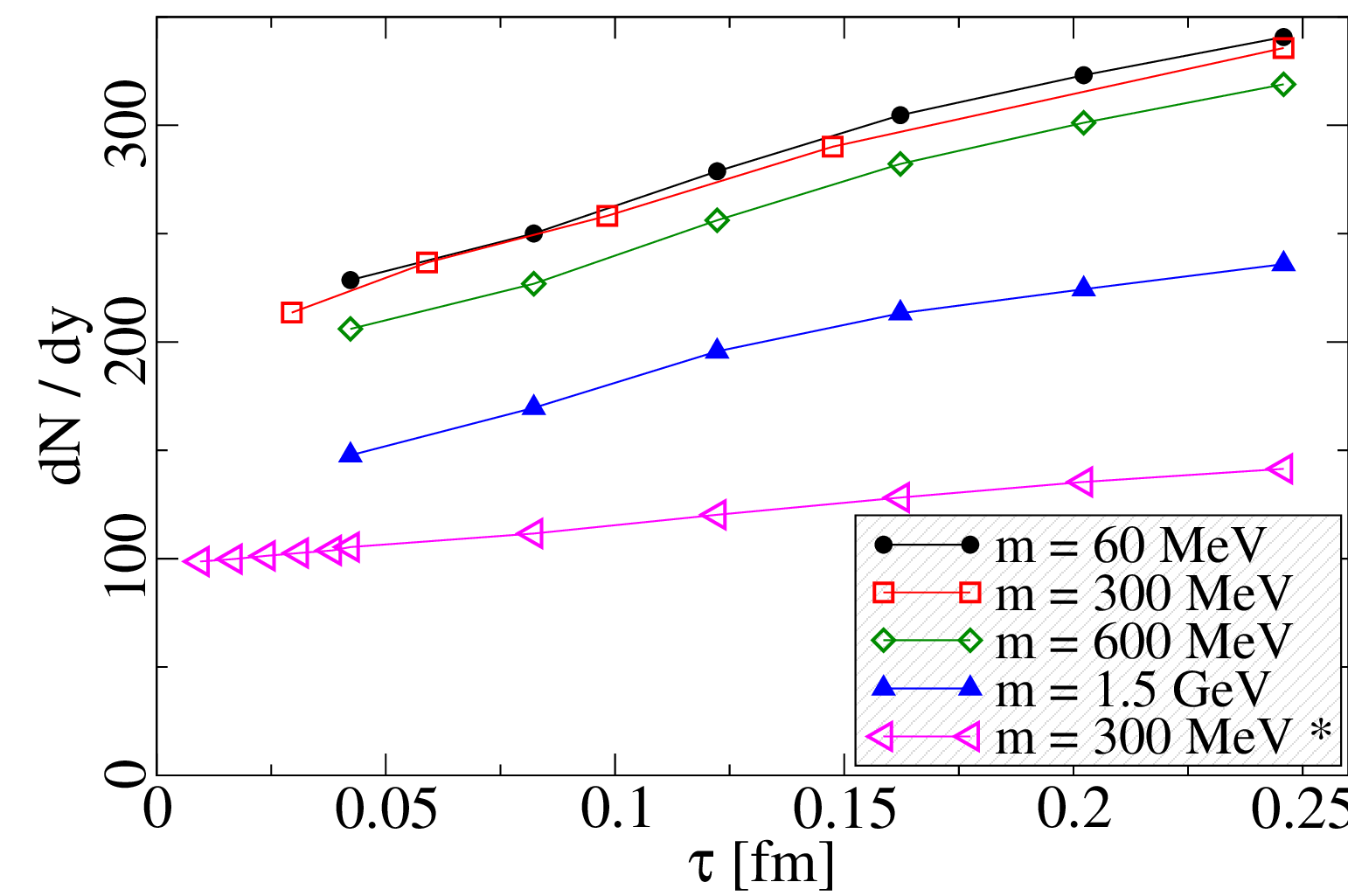}
\end{center}
\vspace{-0.5cm}
\caption{
Dependence on proper time $\tau$ of the number of pairs
of one flavor per unit rapidity $\ud N/\ud y$
for $g^2\mu=2$ GeV and  different quark masses.
The lowest curve  corresponds to $g^2\mu=1 \gev$.
}
\label{fig:taudep}
\end{minipage}
\end{figure}

One starts in the the infinite past $t\to -\infty$ with a negative
energy plane wave solution $\psi(x) = e^{i q \cdot x}v(q)$. The Dirac equation
can then be integrated forward in time analytically to the future light cone
($\tau^2 = 2 x^+x^-=0, \ x^\pm > 0$) because the background field in the 
intermediate region
is a pure gauge. This gives an initial condition,
given explicitly in Eq.~(16) of Ref.~\cite{Gelis:2004jp}, for 
$\psi(\tau=0,z,\xt)$.
Starting from this initial condition we then solve numerically the Dirac equation
for $\tau \ge 0$ using the coordinate system  $\tau,z,\xt$.
The reason for this choice of coordinates is the following.
It not feasible to have degrees of freedom at different energy scales 
($\sqrt{s}$ and $\qs$) present in the same numerical calculation. Thus 
the temporal coordinate is chosen to be $\tau$ in order to include 
the hard sources of the color fields only in the initial condition. 
Although the background field is boost invariant, the longitudinal coordinate
can not be disregarded in this computation because the rapidities of
the quark and the antiquark are correlated. The longitudinal
coordinate is chosen as $z$, not the usual dimensionless 
$\eta = \tanh^{-1} (z/t)$, because the initial condition on the light cone
involves dimensionful longitudinal momentum scales (coming from the four momentum $q$),
and in order to represent them on a spatial lattice at $\tau=0$ a dimensionful coordinate 
is needed. 

The pair production amplitude $M_\tau$ at proper time $\tau$ is then obtained by fixing 
the Coulomb gauge in the transverse plane $\partial_i A_i = 0$ and projecting
the spinor wavefunction $\psi(\tau,z,\xt)$ onto positive
energy states $\phi_p(\tau,z,\xt)  \equiv e^{-ip\cdot x}u(p)$:
\begin{equation}
M_\tau(p,q)\equiv
\int \frac{\tau \ud z \ud^2\xt}{\sqrt{\tau^2+z^2}}
\,\phi_p^\dagger(\tau,z,\xt)\gt\gamma^\tau
\psi_q(\tau,z,\xt).
\label{eq:mtau}
\end{equation}
For times larger than the formation time of the quark pair 
$\tau \gtrsim 1/\sqrt{m^2 + \qt^2}$ this amplitude can be interpreted as the amplitude 
for producing quark antiquark pairs.

\subsection{Results}

 Figure~\ref{fig:ampliy} shows $|M_\tau|^2$
as a function of $\Delta y = y_p - y_q$ integrated over $\pt$ for different $\qt$.
When integrating also over the rapidity difference $\Delta y$ one gets the number of
pairs per unit rapidity as a function of $\tau$,  shown 
in Fig.~\ref{fig:taudep}. It can be seen that the quark production amplitude
reaches a finite value instantaneously and then increases slowly with $\tau$.

\begin{figure}[t]
\begin{minipage}[t]{0.48\textwidth}
\includegraphics[width=0.99\textwidth]{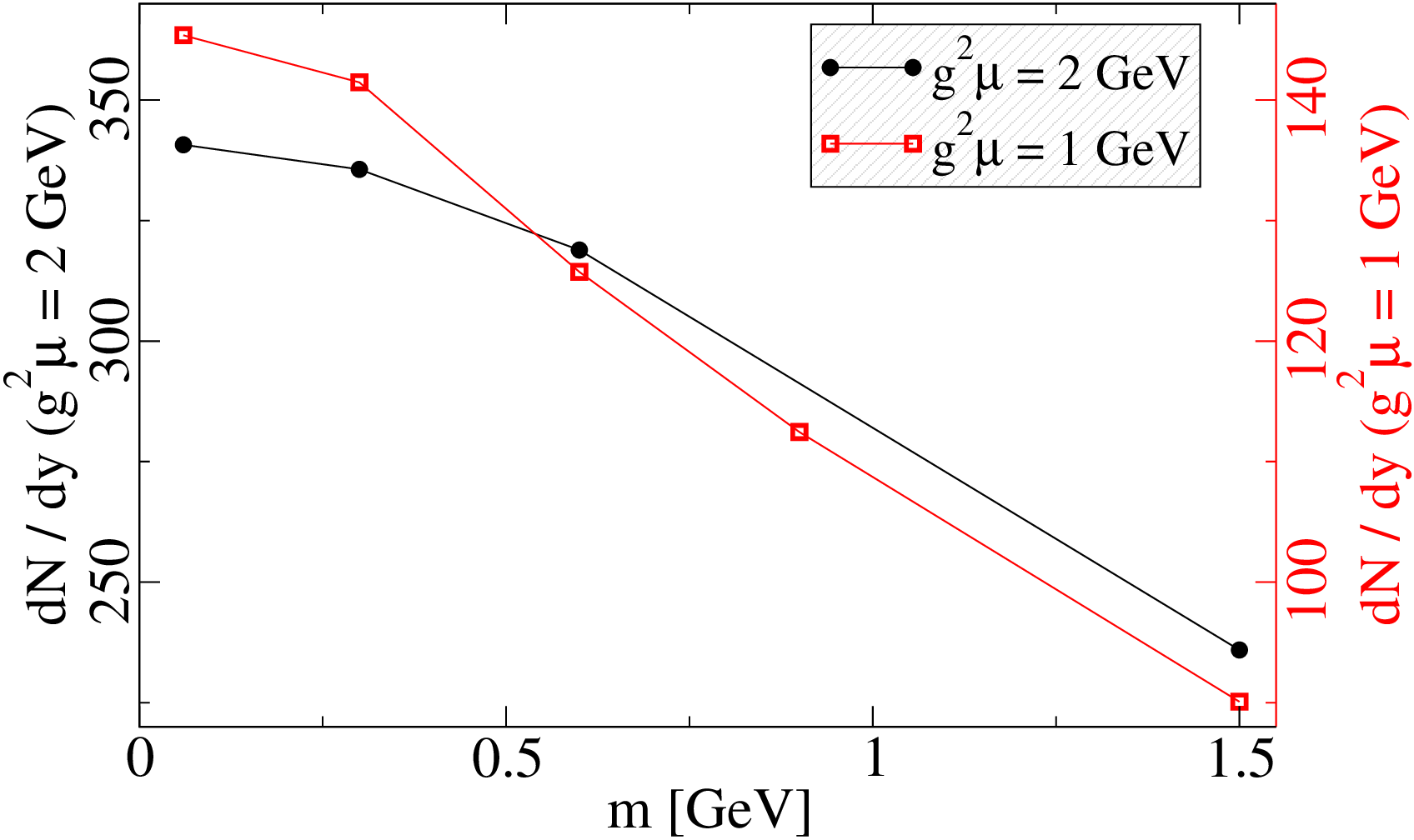}
\vspace{-0.5cm}
\caption{
Dependence of the number of quark pairs
on quark mass at a fixed proper time, $\tau=0.25 \textrm{ fm}$,
and for two values of $g^2\mu$.
}
\label{fig:gsqrmu}
\end{minipage}
\hfill
\begin{minipage}[t]{0.48\textwidth}
\includegraphics[width=0.99\textwidth]{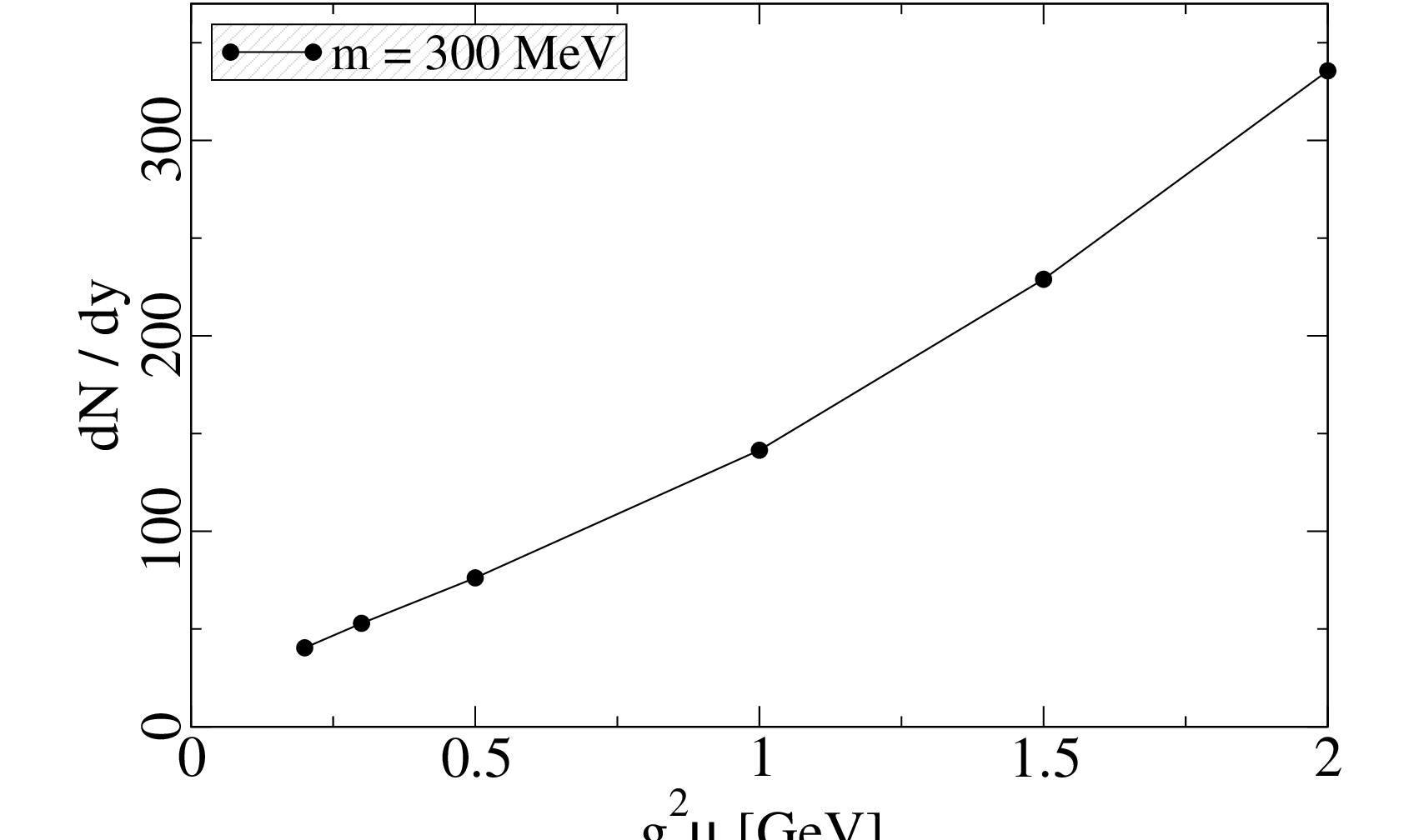}
\vspace{-0.5cm}
\caption{
Dependence of the number of quark pairs on $g^2\mu$ at a
fixed proper time, $\tau=0.25 \textrm{ fm}$, and for quark mass $m=0.3 \textrm{ GeV}$.
} 
\label{fig:mass}
\end{minipage}
\end{figure}

\begin{figure}[t]
\begin{minipage}[t]{0.48\textwidth}
\includegraphics[width=0.99\textwidth]{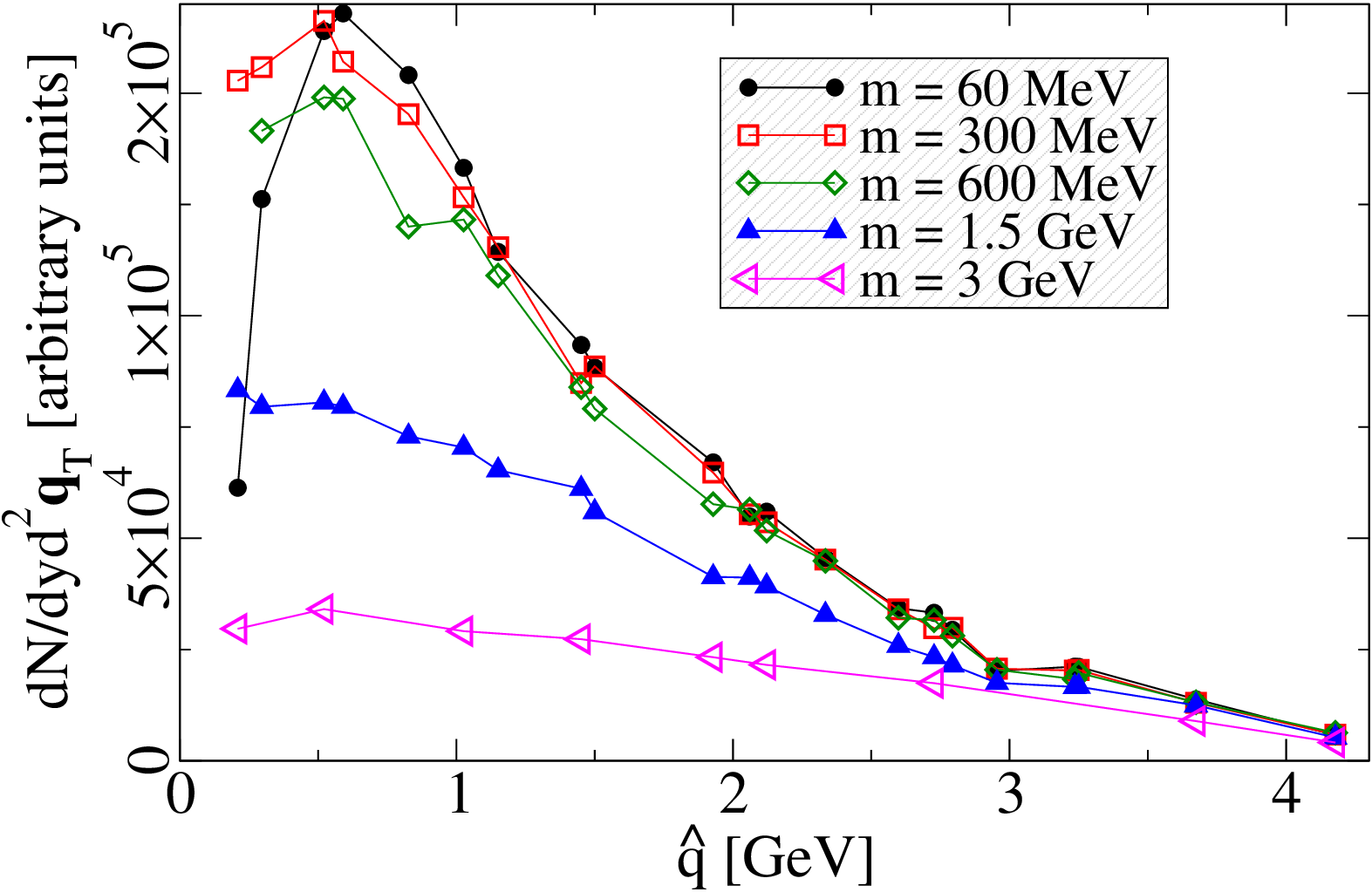}
\vspace{-0.5cm}
\caption{
Transverse momentum spectrum of (anti)quarks for
$g^2\mu=2$ GeV at a fixed proper time, $\tau=0.25$ fm, and
for different quark masses.
}
\label{fig:spectsm}
\end{minipage}
\hfill
\begin{minipage}[t]{0.48\textwidth}
\includegraphics[width=0.99\textwidth]{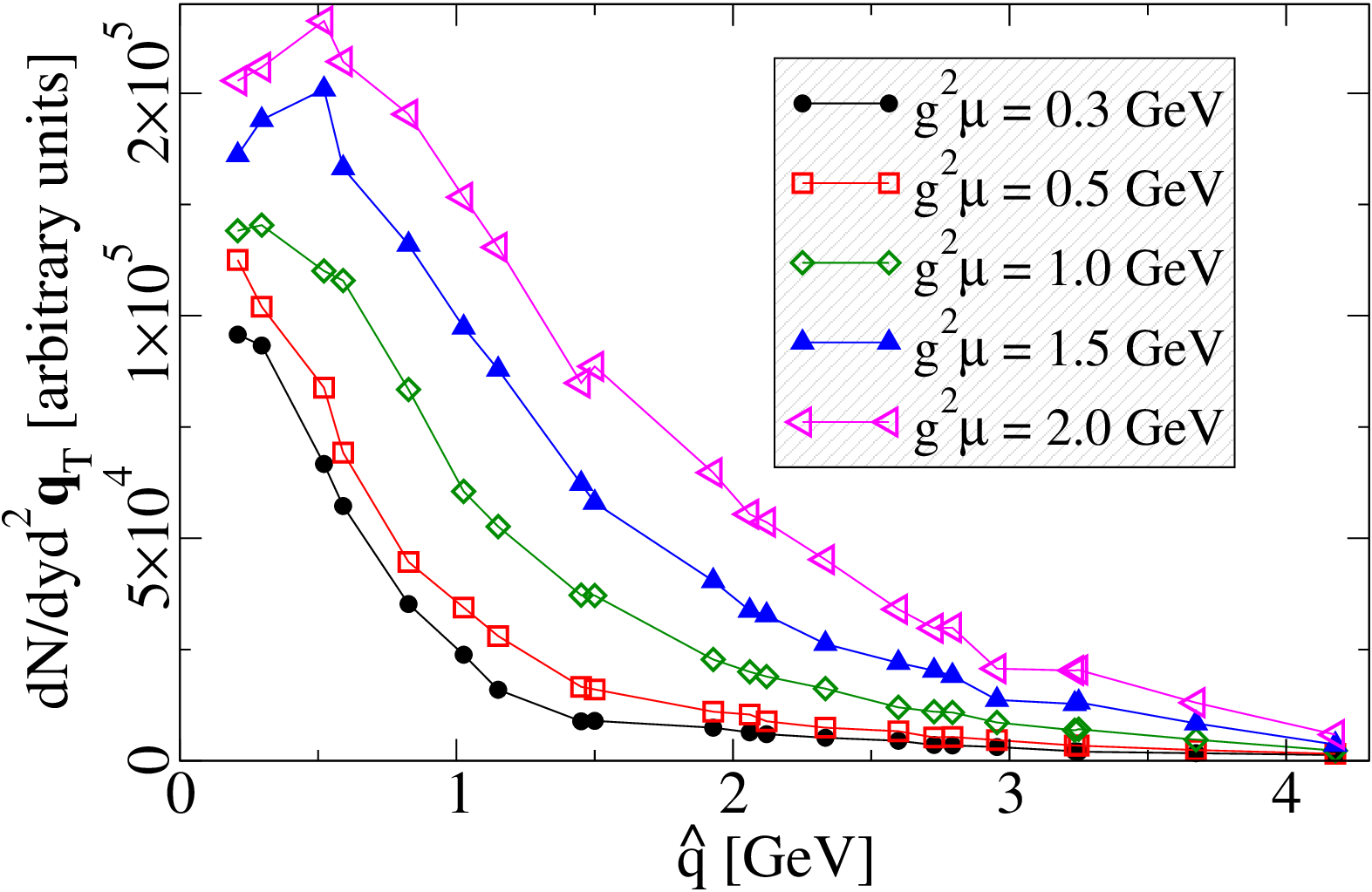}
\vspace{-0.5cm}
\caption{
 Transverse momentum spectrum of (anti)quarks for quark mass
$m=0.3$ GeV and for different $g^2\mu$ at a fixed proper time, $\tau=0.25$ fm.
} 
\label{fig:spectsgsqrmu}
\end{minipage}
\end{figure}

The physical parameters of the calculation are  $g^2\mu$ characterizing the
strength of the background field, the nuclear radius $\ra$ and  the quark mass $m$.
The dependence on $g^2\mu$ and $m$ of the number of pairs at $\tau=0.25 \fm$ is shown in
Figs.~\ref{fig:gsqrmu} and \ref{fig:mass}. The number of pairs 
is seen to increase with $g^2 \mu$, but not as strongly as the 
$(g^2\mu)^2$ dependence predicted by a simple dimensional analysis argument.
The result also decreases with increasing quark mass, but the 
perturbative $1/m^2$-behavior is not reached in our calculation.
The transverse momentum spectra of the (anti)quarks as a 
function of $\qt$ is shown for different quark masses and saturation scales in 
Figs.~\ref{fig:spectsm}  and \ref{fig:spectsgsqrmu}. 

\subsection{The numerical method}
\label{sec:numerics}

Our numerical method is presented in more detail in Refs.~\cite{Gelis:2004jp,Lappi:2005tt}.
The discretization in the transverse plane is straightforward, but 
in the longitudinal direction the $\tau,z$ coordinate system
can easily result in an unstable one. In Ref.~\cite{Gelis:2004jp} a stable
scheme was found by using an implicit method, where at each timestep
one solves a linear system
of equations for the spinor at different points on the longitudinal lattice.

The numerical computation depends on several discretization parameters:
the lattice size and spacing in the longitudinal ($N_z$ and $\ud z$) and the transverse 
($N^2$ and $a$) directions and the timestep $\ud \tau$. To check our numerical method we
studied the dependence on these parameters. We have also tested the numerical method
in the analytically solvable case of zero external field and tested how well our
numerical implementation preserves boost invariance.

\begin{figure}[t]
\begin{minipage}[t]{0.48\textwidth}
\includegraphics[width=0.99\textwidth]{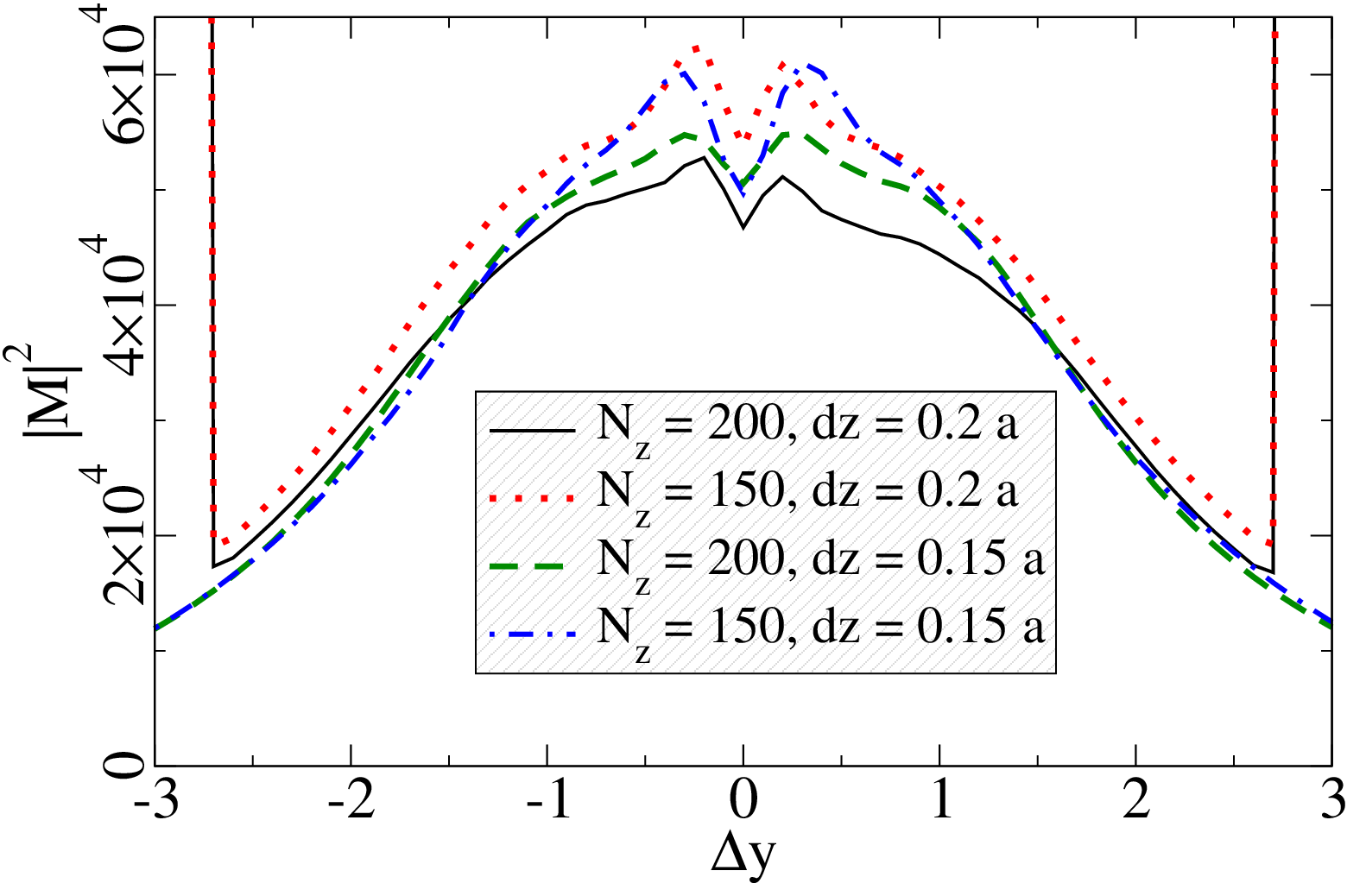}
\vspace{-0.5cm}
\caption{
Amplitude at $\tau= 0.25 \fm$ for $g^2 \mu=2\gev$
for different values of $\ud z$ and $N_z$ . The lattice spacing is measured
in units of $a$, the transverse lattice spacing. 
}
\label{fig:nzdzampliy}
\end{minipage}
\hfill
\begin{minipage}[t]{0.48\textwidth}
\includegraphics[width=0.99\textwidth]{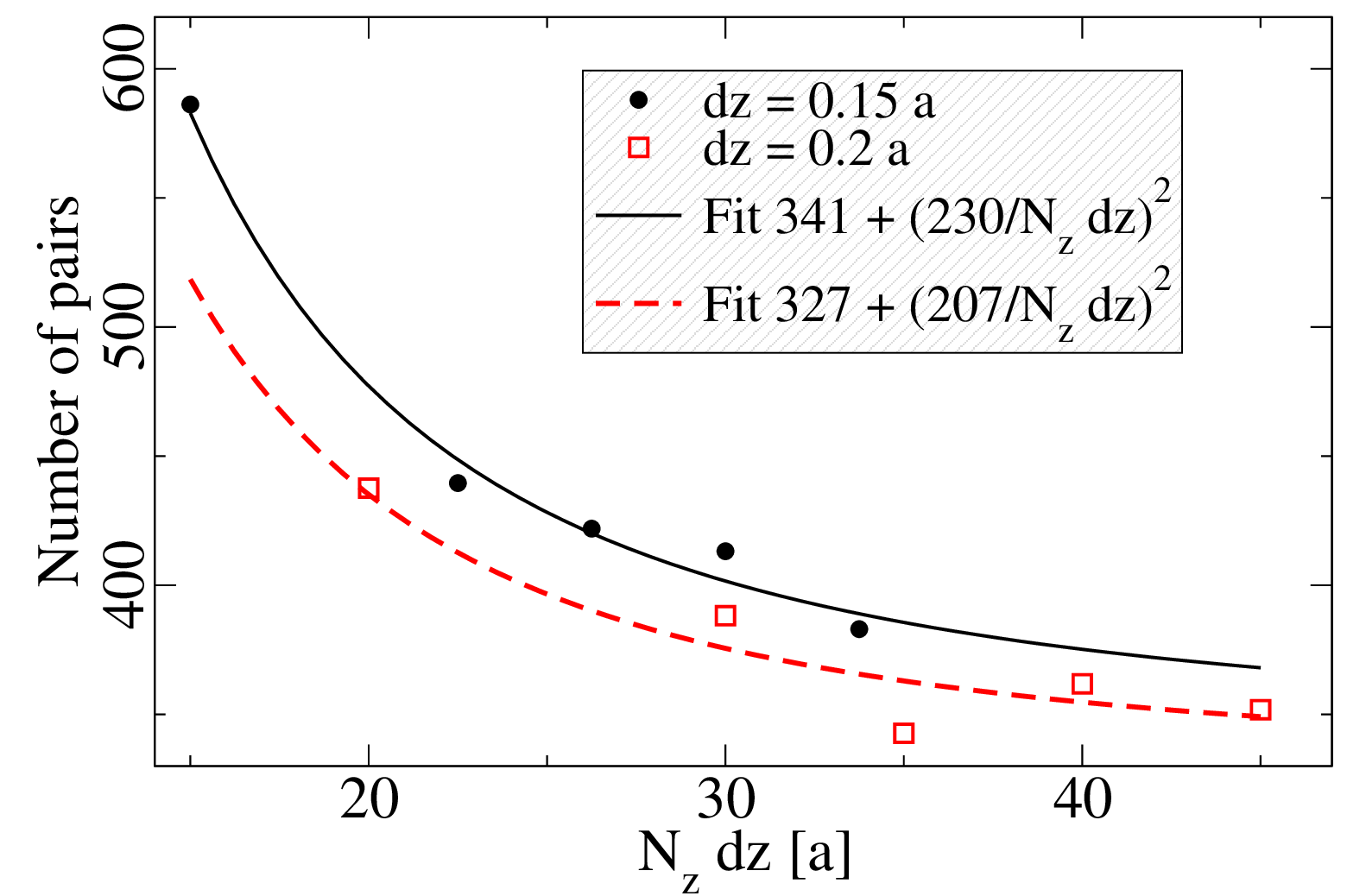}
\vspace{-0.5cm}
\caption{
Number of pairs as a function of (half) the volume of the longitudinal lattice
$N_x \ud z$ for two different values of $\ud z$. The solid lines are fits of the 
form $A + B/(N_z \ud z)^2$. 
} 
\label{fig:nzdz}
\end{minipage}
\end{figure}

\begin{figure}[th]
\begin{minipage}[t]{0.48\textwidth}
\includegraphics[width=0.99\textwidth]{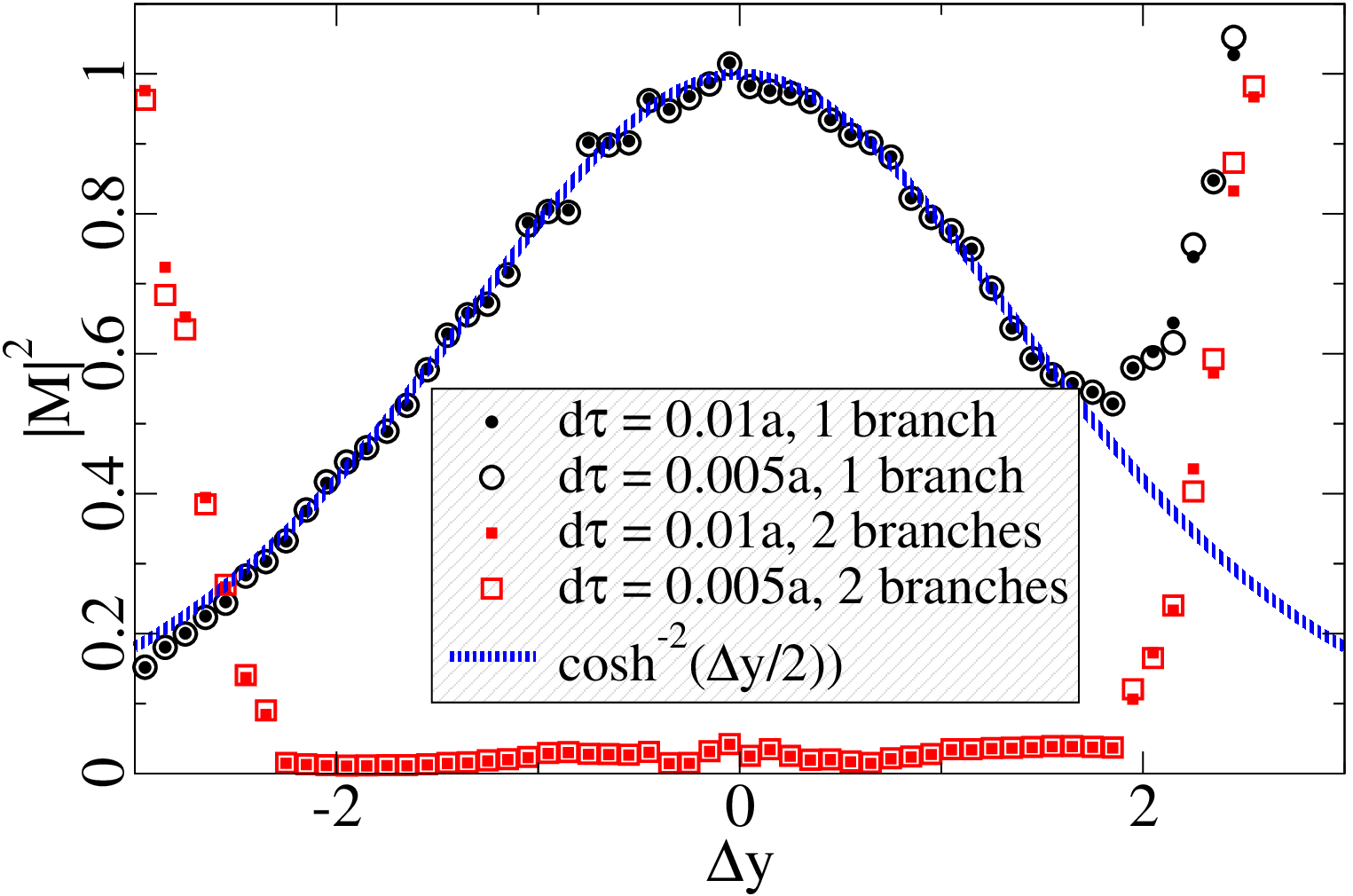}
\vspace{-0.5cm}
\caption{
The amplitude in a zero external color field for two different timesteps $\ud \tau$.
Plotted are the absolute value of the amplitude for only one and both branches 
of the amplitude (see text and Fig.~\ref{fig:spacet}).
}
\label{fig:zerof}
\end{minipage}
\hfill
\begin{minipage}[t]{0.48\textwidth}
\includegraphics[width=0.99\textwidth]{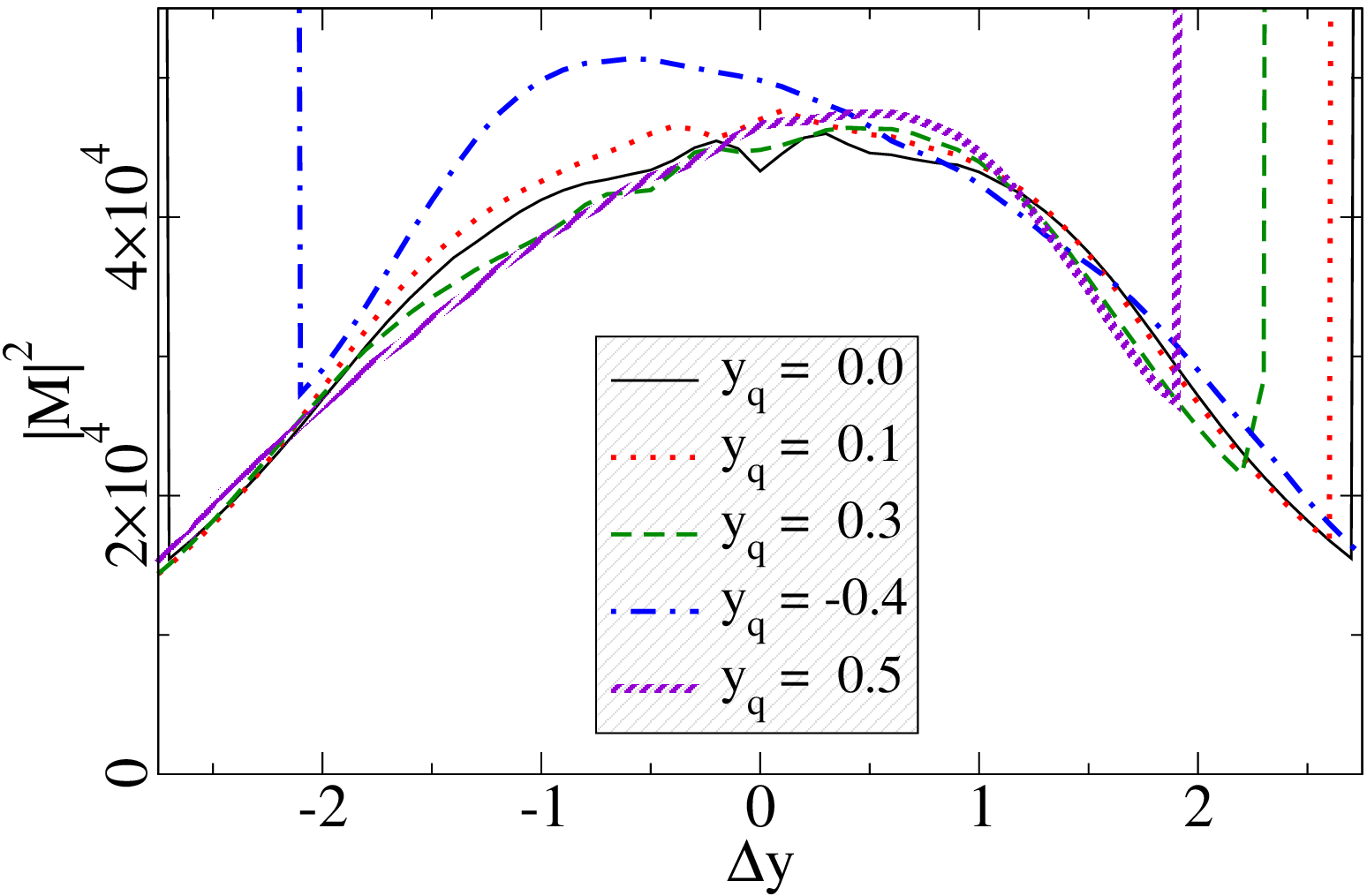}
\vspace{-0.5cm}
\caption{
The amplitude $|M_\tau|^2$ as a function of the rapidity difference $\Delta y$
for different rapidities of the antiquark $y_q$. 
} 
\label{fig:yinit}
\end{minipage}
\end{figure}

The dependence on the longitudinal lattice parameters, $\ud z$ and $N_z$, is studied in 
Figs.~\ref{fig:nzdzampliy} and \ref{fig:nzdz}. Our convention is that the 
longitudinal lattice has points $i=-N_z \dots N_z$, i.e. $2N_z+1$ points and the 
length of the longitudinal lattice is $2N_z\ud z$. The results presented earlier 
in this talk are for $\ud z = 0.2a$ and $N_z = 200$ and have not been
extrapolated to the infinite volume limit  $N_z \ud z \to \infty$. 
A finite $\ud z$ leads to a lattice cutoff in $p^z$ and thus sets a maximum 
for the accessible interval in $\Delta y$.
As can be seen from Fig.~\ref{fig:ampliy} and understood as a consequence of the
initial condition, Eq.~(16) of Ref.~\cite{Gelis:2004jp}, the accessible region 
in $\Delta y$ is smaller for \emph{small} transverse momenta $\qt$.  This cutoff 
in $\Delta y$ is also apparent in Figs.~\ref{fig:nzdzampliy}, \ref{fig:zerof} 
and \ref{fig:yinit}.

We have so far only used transverse lattices of $180^2$ points. The lattice 
momenta can be represented as  $\qt=(q_x,q_y)$ with $q_{x,y} =-89 \dots 90$, of which the
$q_{x,y} = -45 \dots 45$ are non-doubler modes.
We explicitly leave out the doubler modes both in the initial condition ($\qt$ modes) and
in the projection to the positive energy state ($\pt$ modes). This limits the 
volume of the transverse momentum space to $1/4$ of the bosonic case. Also the spectrum
of quarks (see Figs.~\ref{fig:spectsm} and \ref{fig:spectsgsqrmu}) decreases
so slowly for large transverse momenta that some dependence on the $1/a$ lattice 
cutoff is expected. Further computations with larger transverse lattices are still needed to 
study this issue.

The memory requirement for a $400\times180^2$--lattice with $4\times (\nc=3)$
complex components in one spinor is $1.2 \textrm{ GB}$ in single precision. 
This can still be achieved on one processor, but for larger lattices a parallelized
version of the program will have to be used.
The numerical computations have been done on the ``ametisti'' cluster,
a $66\times2$ processor AMD Opteron Linux cluster at the University of Helsinki,
using over $10^{17}$ flop of CPU-time so far.

As explained in Fig.~\ref{fig:spacet}, the amplitude is a sum of two terms. For a zero
external color field these two terms, with an absolute value
$|M_\tau| = 1/\cosh (\Delta y/2)$ have an opposite sign and cancel each other. Comparing
the analytically and numerically computed amplitudes is a nontrivial test of the numerical 
computation. Figure~\ref{fig:zerof} shows how the analytical result 
for one branch and the cancellation when both branches are included are reproduced 
by our numerical method for one value of $\qt$ ($\qt=(5,5)$ on a $150^2$ lattice). 

Due to the boost invariance of the background field the amplitude $M_\tau$ should not depend 
on the rapidities $y_q$ and $y_p$ separately, but only on the difference 
$\Delta y \equiv y_p-y_q$. Because the calculation is done using $z$, not rapidity, 
as the longitudinal 
coordinate, verifying the boost invariance of the resulting amplitude is a nontrivial test
of the numerical method. Figure~\ref{fig:yinit} shows that the number of pairs is, 
taking into account numerical inaccuracies, not
affected by the choice of $y_q$.

\section{Discussion}

\begin{figure}
\includegraphics[width=0.43\textwidth]{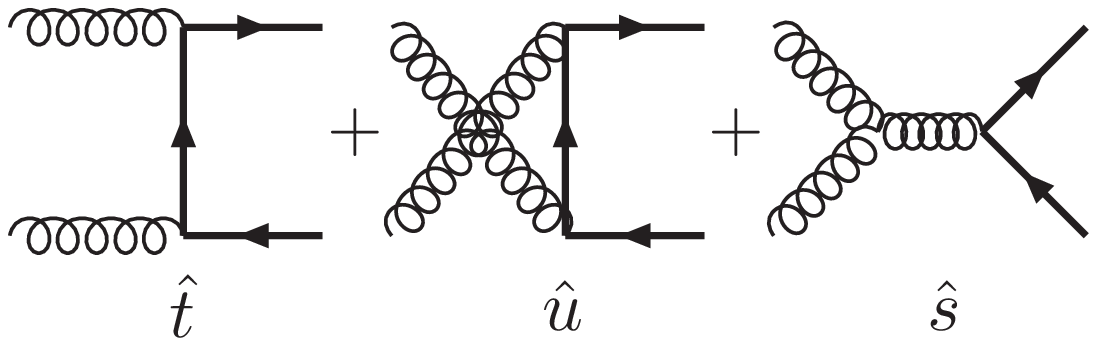}
\hfill
\includegraphics[width=0.57\textwidth]{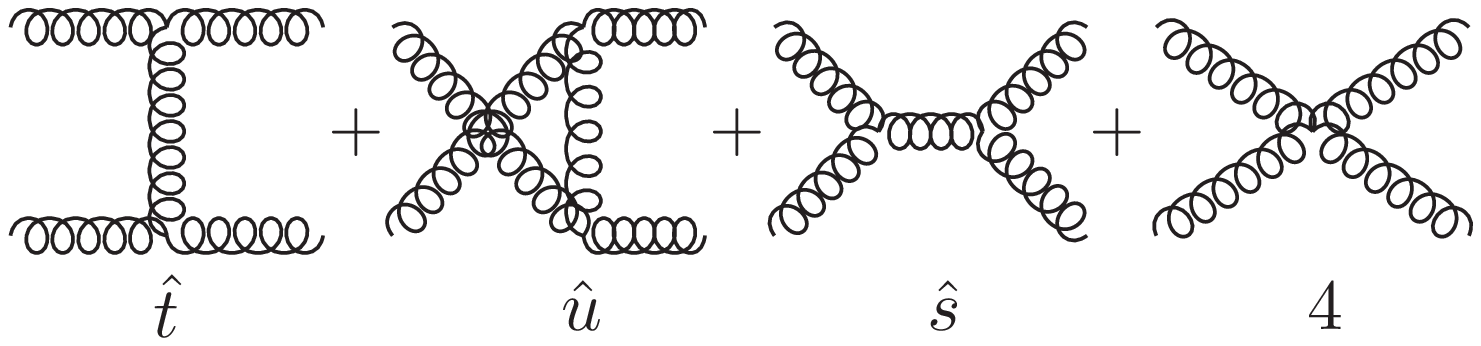}
\vspace{-1cm}
\caption{
The \mbox{2 $\to$ 2}-processes for producing quark pairs and gluons from an initial 
gluon distribution. }
\label{fig:diags}
\end{figure}

It has conventionally been assumed that the initial state of a
heavy ion collision is dominated by gluons. This is the
result e.g. when both quarks and gluons are produced in
$2 \to 2$ collisions of collinear partons \cite{Eskola:1999fc}.
Indeed, when comparing the cross sections of the $2 \to 2$ processes 
for quark pair production and gluon production (see Fig.~\ref{fig:diags}), 
the diagrams for quark pair production are suppressed by a factor of 
$\sim 210 = 7 \times 30$, of which the factor 7 is due to color factors.

In the color glass condensate framework the picture is quite different. Whereas the
quark pairs are produced in a $2 \to 2$ process similarly to collinear factorisation, 
gluons are produced in what reduces in the perturbative limit to a $2 \to 1$ process
with a smaller power of the coupling $g$. It is therefore less straightforward to compare
the two. Strictly perturbatively quark pair production is 
of higher order in $g$. But the phenomenologically relevant value is \mbox{$g\approx 2$}, 
and thus higher orders in $g$ are not  suppressed by much when looking at the 
actual numbers. One can then legitimately ask whether one should, for consistency,
also include quantum corrections (higher orders in $g$) in the computation of gluon 
production.

Brushing aside these remarks and boldly taking our numerical results at their face
value would lead to the following scenario. It was earlier assumed that
the initial state of a heavy ion collision
is dominated by gluons. If the subsequent evolution of the system 
conserves entropy and thus, approximately, multiplicity
this would mean $\sim 1000$  gluons in a unit of rapidity.
In the classical field model this corresponds \cite{Lappi:2003bi} to 
$g^2\mu \approx 2 \gev$. Our results seems to point to a rather large number 
of quark pairs present already in the initial state. One could envisage a scenario
where for $g^2\mu \approx 1.3 \gev$ these 1000 particles could consist of
$\gtrsim 400$ gluons, $\gtrsim 300$ quarks and $\gtrsim 300$ antiquarks
(take the lowest curve from Fig.~\ref{fig:taudep} and multiply by $\nf=3$).
This would be close to the thermal ratio of $N_g/N_q = 64/(21 \nf)$.

\section{Conclusions}

We have calculated 
quark pair production from classical background field of McLerran-Venugopalan 
model by solving the 3+1--dimensional Dirac equation numerically 
in this classical background field. We find that 
number of quarks produced is large, pointing to a possible fast
chemical equilibration of the system.
The mass dependence of our result surprisingly weak
and we are not yet able to make any conclusions on heavy quarks until
studying the numerical issues involved.

\subsection*{Acknowledgements}
T.L. was supported by the Finnish Cultural
Foundation.  This research has also been supported by the Academy of
Finland, contract 77744. We thank  R. Venugopalan,
H. Fujii, K. J. Eskola, B. Mueller and D. Kharzeev for discussions.

\bibliographystyle{JHEP-2mod}
\bibliography{spires}

\vfill\eject
\end{document}